\documentclass[usenatbib,preprint]{mn2e}

\usepackage{graphicx}
\usepackage {natbib,aas_macros}
\usepackage {mediabb}
\usepackage {bm}
\usepackage {amsmath,amssymb}

\newcommand{\msun}{\thinspace M_\odot} 
\newcommand{\gcm}{~{\rm g~cm}^{-3} }
\newcommand{\St}{~{\rm St} }

\newcommand{\vel}{\mathbf{v}}

\newcommand{\ms}{~{\rm m} ~{\rm s}^{-1} } 
\newcommand{\kms}{~{\rm km} ~{\rm s}^{-1} }

\newcommand{\mum}{{\rm \mu} {\rm m} }
\newcommand{\mm}{{\rm mm}}
\newcommand{\cm}{{\rm cm}}

\newcommand{\msunyear}{\thinspace M_\odot~{\rm yr}^{-1}}

\newcommand{\dv}{\Delta \vel}

\newcommand{\tstop}{t_{\rm stop}}
\newcommand{\tgrowth}{t_{\rm growth}}

\title{``Ash-fall" induced by molecular outflow in protostar evolution}

\author[Tsukamoto et al]{
Yusuke Tsukamoto$^{1}$, Masahiro N. Machida$^{2}$, and  Shu-ichiro Inutsuka$^{3}$ \\
$^1$Graduate Schools of Science and Engineering, Kagoshima University, Kagoshima, Japan  \\
$^2$Department of Earth and Planetary Sciences, Kyushu University, Fukuoka, Japan \\
$^3$Department of Physics, Nagoya University, Aichi, Japan  \\
}

\begin{document}
\maketitle

\begin{abstract}
  Dust growth and its associated dynamics play key roles in the first phase of planet formation in young stellar objects (YSOs).
  Observations have detected signs of dust growth in very young protoplanetary disks.
  Furthermore, signs of planet formation, gaps in the disk at a distance of
  several 10 astronomical units (AU) from the central protostar are also reported.
  From a theoretical point of view, however, it is not clear how planet form at the outer region of a disk
  despite the difficulty due to rapid inward drift of dust so called radial drift barrier.
  Here, on the basis of three-dimensional magneto-hydrodynamical simulations of disk evolution with the dust growth,
  we propose a mechanism named ``ash-fall" phenomenon induced by powerful molecular outflow driven by magnetic field which may circumvent the radial drift barrier.
  We found that the large dust which grows to a size of $\sim \cm$ in the inner region of a disk is entrained by an outflow from the disk.
  Then large dust decoupled from gas is ejected from the outflow due to centrifugal force, enriching the grown dust in the envelope and is eventually fall onto the outer edge of the disk.
  The overall process is similar to behaviour of ash-fall from volcanic eruptions.
  In the ash-fall phenomenon, the Stokes number of dust increases by reaccreting to the less dense disk outer edge.
  This may make the dust grains overcome the radial drift barrier. Consequently, the ash-fall phenomenon
  can provide a crucial assist for making the formation of the planetesimals in outer region of the disk possible, 
  and hence the formation of wide-orbit planets and the formation of the gaps.
\end{abstract}


\section{Introduction}
\label{intro}
The physics of dust growth and related dynamics during protostar evolution has attracted a great deal of attention.
Dust growth is the first step in planet formation, and knowing when and where dust growth begins
is an essential part for a unified understanding of star and planet formation.
Recent observations of dust thermal emission have shown that dust growth begins in early phase of protostar evolution.
For example, the observed decrease in the dust spectral index can be interpreted as a sign of dust growth
\citep{2009ApJ...696..841K,2014A&A...567A..32M,2015ApJ...813...41P,2016A&A...588A..53T,2019ApJ...883...71C}.
In addition, the variation of the dust polarization pattern with different wavelength, as observed in some sources in e.g., HL Tau, 
may be due to self-scattering of grown dust \citep{2016ApJ...820...54K}.

Another rather indirect observational indication of dust growth and planet formation in the early phase of protostar evolution is 
the gaps at several tens of AU in protoplanetary disks \citep{2015ApJ...808L...3A,2017A&A...600A..72F,2018ApJ...869L..41A,2018A&A...610A..24F,2018ApJ...869...17L}.
The widely accepted hypothesis to explain the gaps is that the planets already formed in the disk generate the gaps through their gravitational torque
\citep{2015MNRAS.453L..73D,2015ApJ...806L..15K}.
The hypothesis assumes planet formation in the outer region of the disk at a distance of several tens AU from the central protostar prior to the gap formation.
However, this contradicts the prediction of the standard theory of dust dynamics and evolution in the protostar system; as dust grows in the early phase of disk evolution
and its Stokes number increases to $\St\sim 1$, the dust begins to drift inwards due to headwind of disk gas,
thereby suppressing planetesimal and planet formation, especially at a distance of 
several tens of AU from the central protostar \citep{2012ApJ...752..106O,2017ApJ...839...16C},
the effect of which is known as the radial drift barrier \citep{1977MNRAS.180...57W}.
Therefore, some unknown mechanism which counter or circumvents the radial drift in the outer region is necessary to corroborate the standard theory about the gap formation.

Previous studies which investigated dust growth in the early stages of protostar evolution have been based on one-zone approximations or one-dimensional simulations
of isolated disk \citep{2012ApJ...752..106O,2012A&A...539A.148B,2013MNRAS.434L..70H}
or have ignored the effect of magnetic fields \citep{2018A&A...614A..98V}.
In other words, none of them considered how the dust grows and dynamically evolves in three-dimensional space  under the influence of various gas dynamics in real protostellar system, such as outflow driving induced by magnetic fields.

  Recently \citet{2020A&A...641A.112L} investigated
  dust dynamics in the cloud core collapse with three-dimensional MHD simulation
  with dust size-distribution,
  and suggested that the dust with a size of $\gtrsim 100 \mum$
  existing in the molecular cloud core radially drifts during its collapse,
  and settles in the newly born disk and enhances its dust-to-gas mass ratio.
  However, the dust growth is not calculated in their simulations.

In this paper, on the basis of three-dimensional magneto-hydrodynamical simulations of dust-gas mixture which consider the dust growth, 
we propose a mechanism, the ``ash-fall" phenomenon induced by molecular outflow driven by magnetic field in protostar evolution
that can circumvent the radial drift barrier,

The paper is organized as follow.
In \S 2, we describe the method and initial condition of our simulation.
Then in \S 3, we describe the simulation results.
Finally, our results are discussed in \S 4.

\section{Methods and initial condition}
\subsection{Numerical methods}
We solve two-fluid magneto-hydrodynamics (MHD) equations for dust-gas mixture.
The detail of the governing equations and numerical scheme are described in our previous paper \citep{2021ApJ...913..148T}.
In \citet{2021ApJ...913..148T}, we also showed that treating charged dust and neutral dust
as a single fluid is justified for investigating the dynamics of the dust with size of $a_d\gtrsim 10 \mum$, which is the focus of this paper.

In addition to the governing equations presented in \citet{2021ApJ...913..148T},
we further solve the equation for the dust-size evolution with single-size approximation \citep[e.g.,][]{2016A&A...589A..15S,2017ApJ...838..151T} as follows.
The equation for the time evolution of the dust-size $a_d$ is given by
\begin{eqnarray}
  \label{dadt}
  \frac{D_d a_d}{Dt}=A_{\rm gain/loss} \frac{a_d}{3 t_{\rm growth}}.
\end{eqnarray}
where $t_{\rm growth}=1/(\pi a_d^2 n_d \Delta v_{\rm dust})$, $n_d$ is the dust number density,
$\Delta v_{\rm dust}$ is the collision velocity between the dust grains,
$D_d/Dt=\partial/\partial t +v_d \cdot \nabla$, $v_d$ is the (mean) dust velocity,
and
\begin{eqnarray}
  \label{collision_factor}
A_{\rm gain/loss}={\rm min}(1,-\frac{\ln(\Delta v_{\rm dust}/\Delta v_{\rm frag})}{\ln 5}),
\end{eqnarray}
is a factor to model the collisional mass gain and loss \citep{2016ApJ...821...82O}.
The dust gains and loses a mass when $\Delta v_{\rm dust}  <\Delta v_{\rm frag}$ and 
$\Delta v_{\rm dust}  > \Delta v_{\rm frag}$, respectively,  through dust collisions.
To evaluate the dust relative velocity between the dust $\Delta v_{\rm dust}$, we consider
the sub-grid scale turbulence and Brownian motion,
and relative velocity is given as $\Delta v_{\rm dust}=\Delta v_{\rm turb}+ \Delta v_B$  (for details, see Appendix A).

We model equation \ref{collision_factor} so that it reproduces the
results of the collision simulations of dust aggregates \citep{2013A&A...559A..62W}.
In this study we set $\Delta v_{\rm frag}=30 \ms$.
We can rewrite $D_d/Dt$ as
\begin{eqnarray}
  \frac{D_d}{Dt}&=&\frac{\partial}{\partial t} + \vel_d \cdot \nabla  \nonumber \\
  &=&\frac{\partial}{\partial t} + \vel\cdot \nabla -\frac{\rho_g}{\rho}(\vel_g-\vel_d)\cdot \nabla \nonumber \\
  &=&\frac{D}{D t} + (1-\epsilon) \dv \cdot \nabla,
\end{eqnarray}
where $\vel$ is the barycentric velocity of the dust-gas mixture,
$\rho_g$ is the gas density, 
$\vel_g$ is the gas velocity,
$\rho=\rho_g+\rho_d$ is the total density,
$\rho_d$ is the dust density,
$\epsilon=\rho_d/\rho$, and 
$\dv=\vel_d-\vel_g$.
Using this relation, we can rewrite the equation (\ref{dadt}),
\begin{eqnarray}
  \frac{D a_d}{Dt}=A_{\rm gain/loss} \frac{a_d}{3 \tgrowth}-(1-\epsilon)\dv \cdot \nabla a_d.
\end{eqnarray}
The SPH discretization form of this equation is given as
\begin{eqnarray}
  \frac{D a_{d,i}}{Dt}&=&A_{\rm gain/loss} \frac{a_{d, i}}{3 \tgrowth}  \nonumber \\
  &-&(1-\epsilon_i) \sum_j m_j (a_{d,j}-a_{d,i})\frac{\dv_i \cdot \nabla W_{ij}(h_i)}{\Omega_i \rho_i} \nonumber \\
  &+& \frac{\alpha_{\dv} v_{\rm sig,\dv}}{\bar{\rho}}(a_{d,i}-a_{d,j}) \mathbf{e}_{ij}\cdot \bar{\nabla W_{ij}},
\end{eqnarray}
where the last term is an artificial viscosity and
$v_{\rm sig,\dv}=\frac{1}{2}(c_{s,i}+c_{s,j})+3 \beta_{\rm visc} {\rm min}(0,(\dv_i-\dv_j)\cdot \mathbf{e}_{ij})$ is the signal velocity where $c_{s, i}$ is the sound velocity of
$i$ th particle and $\mathbf{e}_{ij}=(\mathbf{r}_i-\mathbf{r}_j)/|\mathbf{r}_i-\mathbf{r}_j|$. We choose $\beta_{\rm visc}=2$.

The timestep for dust advection is chosen to be
\begin{eqnarray}
dt_i=\frac{h_i}{(1-\epsilon_i)\dv_i},
\end{eqnarray}
where $h_i$ is the smoothing length of $i$ th particle.

During the simulations, $\epsilon_i$ of a few particles are found to become negative (in particular in the outflow regions) with $|\epsilon_i|\lesssim 10^{-7}$.
When it happens, we correct $\epsilon_i$ to be $|\epsilon_i|$ and maintain it positive.
We confirm that the error of the total dust mass introduced with this correction is negligible 
because the absolute value of the negative $\epsilon_i$,  as well as the number of the particles with $\epsilon_i<0$, is very small.

Our numerical simulations consider non-ideal MHD effects, including the Ohmic and ambipolar
diffusions, but ignore the  Hall effect.
For the resistivity model, we adopt the resistivity table with $a_d=0.035 \mum$ presented in
\citet{2020ApJ...896..158T}. Thus, the dust size for dust dynamics and that for the resistivity table
are not consistent with each other in our present simulations.

A sink particle was dynamically introduced when the density exceeds $\rho_{\rm sink}= 10^{-12} \gcm$.
The sink particle absorbs SPH particles with $\rho>\rho_{\rm sink}$ within $r_{\rm sink}<1$ AU.

\subsection{Initial conditions}
The initial condition of the cloud core is as follows.
We adopt the density-enhanced Bonnor-Ebert sphere surrounded by medium with a steep density
profile of $\rho \propto r^{-4}$ as the initial condition, which is
\begin{eqnarray}
  \rho(r)=\begin{cases}
  \rho_0 \xi_{\rm BE}(r/a) ~{\rm for} ~ r < R_c \\
  \rho_0 \xi_{\rm BE}(R_c/a)(\frac{r}{R_c})^{-4} ~{\rm for} ~ R_c < r < 5 R_c,
  \end{cases}
\end{eqnarray}
with
\begin{eqnarray}
  a=c_{\rm s, iso} \left( \frac{f}{4 \pi G \rho_0} \right)^{1/2},
\end{eqnarray}
where $\xi_{\rm BE}$ is a non-dimensional density profile of the critical Bonnor-Ebert sphere.
The steep density profile in $r>R_c$ is introduced to reduce the particle number and unnecessary computational costs.
$f$ is a numerical factor related to the strength of the gravity, and $R_c=6.45 a$ is the radius of the cloud core.
In this study, we adopt $\rho_0=7.3\times 10^{-18} \gcm$,
$\rho_0/\rho(R_c)=14$, and $f=2.1$. Then, the radius of the core
is $R_c=4.8\times 10^3$ AU and the enclosed mass within $R_c$ is $M_c=1 \msun$.
We adopt an angular velocity  profile of $\Omega(d)=\Omega_0/(\exp(10(d/(1.5 R_c))-1)+1)$
with $d=\sqrt{x^2+y^2}$ and  $\Omega_0=2.3\times 10^{-13} {\rm s^{-1}}$.
We adopt a constant magnetic field $(B_x,B_y,B_z)=(0,0,83 \mu G)$.
The parameter $\alpha_{\rm therm}$ ($\equiv E_{\rm therm}/E_{\rm grav}$) is $0.4$, where
$E_{\rm therm}$ and $E_{\rm grav}$ are the thermal and gravitational energies of the central core
(without surrounding medium), respectively.
The ratio of the rotational to gravitational energies $\beta_{\rm rot}$ within
the core is $\beta_{\rm rot}$ ($\equiv E_{\rm rot}/E_{\rm grav}$) $=0.03$,
where $E_{\rm rot}$ is the rotational energy of the core.
The mass-to-flux ratio of the core is  $\mu/\mu_{\rm crit}=3$.
We resolve 1 $\msun$ with $3\times 10^6$ SPH particles. 
We adopt a dust density profile of $\rho_d(r)=f_{dg} \rho_g(r)/(\exp(10(r/(1.5 R_c))-1)+1)$
where $f_{dg}=10^{-2}$ is the dust-to-gas mass ratio.
The dust density profile  has the same shape
with the gas density profile in $r\lesssim 1.5 R_c$ but is truncated at $r \geq 1.5 R_c$. The initial dust size is $a_d=0.1 \mum$.

\section{Results}
\subsection{Time evolution}
\begin{figure*}
  \hspace{-16 mm}
\includegraphics[width=190mm]{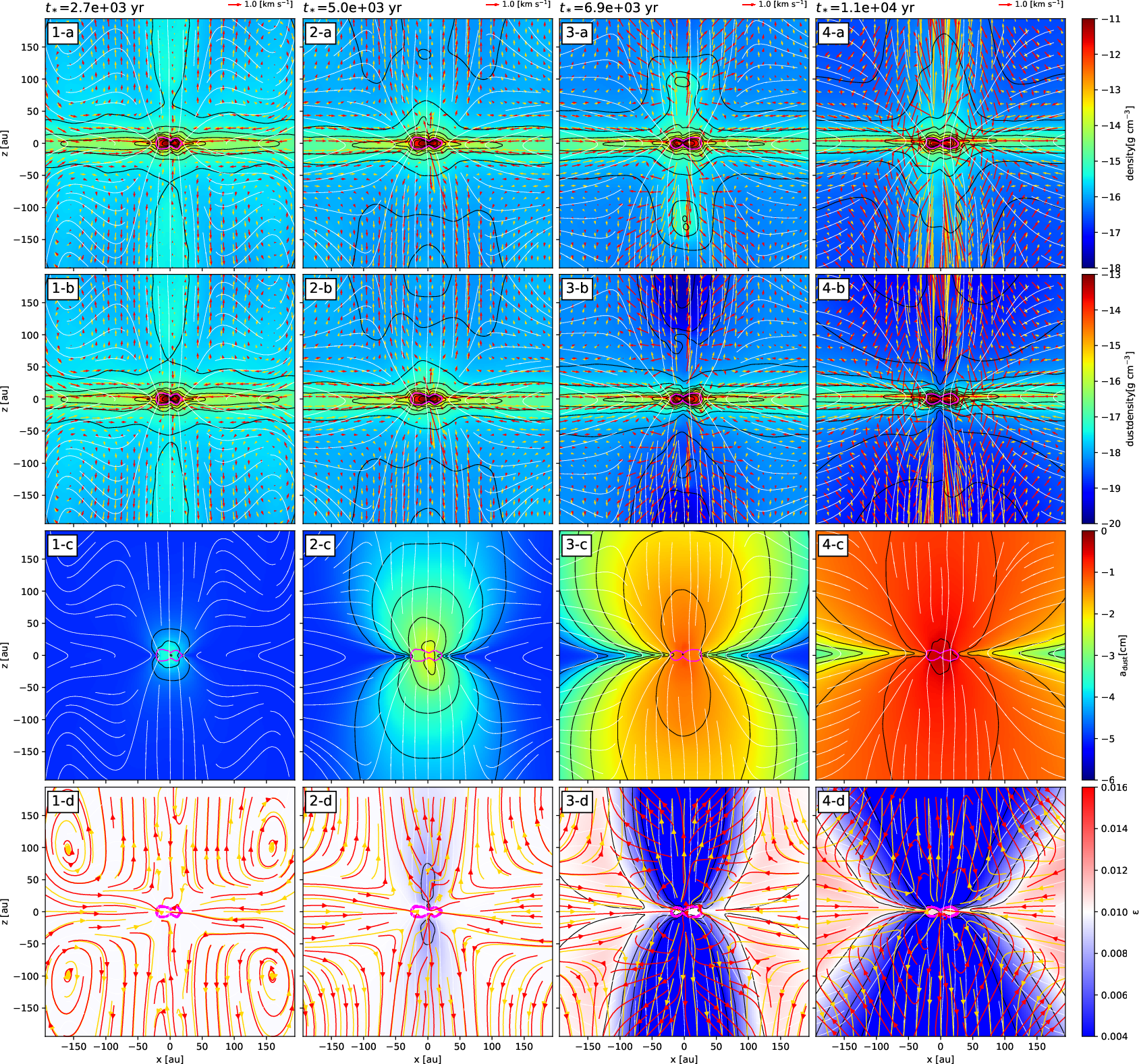}
\caption{Time evolution of the (panels a) gas density, (b) dust density, (c) dust size, and  (d) dust-to-gas mass ratio on the $x$-$z$ plane
  where the $x$- and $z$- axis are perpendicular and parallel to the rotation axis of parent core, respectively.
  Red and orange arrows show the velocity field of the dust and gas, respectively, on the $x$-$z$ plane.
  Red and orange lines show their respective stream-lines.
  White lines show the magnetic field.
  Black lines are the contour of the quantity of each panel.
  The contour levels are (panels a) $\rho_g=10^{-18},10^{-17.5},\cdots,10^{-11}\gcm$, (b) $\rho_d=10^{-20},10^{-19.5},\cdots,10^{-13}\gcm$,
  (c) $a_d=10^{-6},10^{-5.5},\cdots,10^{0}\cm$, and (d) $\epsilon=4\times 10^{-3},\cdots, 1.6 \times 10^{-2}$.
  Magenta lines are the contour of $\rho_g=10^{-13} \gcm$ within which is considered a disk.
}
\label{fig1}
\end{figure*}

We simulated the evolution of a protostar, a protoplanetary disk, and a outflow until $1.2 \times 10^4$ years after the epoch
of protostar formation.
Figure \ref{fig1} shows time evolution at four epochs of the gas density, dust density, dust size, and dust-to-gas mass ratio on the $x$-$z$ plane where
the $x$- and $z$-axes are perpendicular and  parallel to the rotation axis of parent core, respectively.

At the epoch $t_*=3\times 10^3$ yr (panels 1-a to 1-d in Figure \ref{fig1}) where $t_*=0$ corresponds to the formation epoch of protostar,
the bipolar outflow has been formed and gas and dust are blown out from the central region.
At this stage, however, dust has not grown significantly with maximum size of $\lesssim 10 \mum$, which is found at the centre of the system.
The grown dust is distributed only in vicinity of the disk (panel 1-c).
Given the small maximum size of the dust, the dust and gas should be still fully coupled,
which is supported by the facts that their density structures are identical (panels 1-a and 1-b) and that the dust-to-gas mass ratio remains at the initial value of $0.01$ (panel 1-d).

At $t_*=5\times 10^3$ yr (panels 2-a to 2-d), the maximum size of the dust
reaches $\sim 100 \mum$ at the centre, and dust with size of  $a_d \sim 10 \mum$
begins to be entrained by the outflow (panel 2-c).
Panels 2-a, b and d indicate that the dust and gas begin to
decouple in the outflow and dust spreads out of the outflow.
As a result, the dust-to-gas mass ratio begins to decrease inside the outflow (panel 2-d).

At $t_*=7\times 10^3$ yr (panels 3-a to 3-d),
the dust with a size of $a_d \sim 100 \mum$ is distributed over a large area of a few 100 AU across.
The stream-lines and arrows indicate that the dust moves parallel to the $x$-axis inside the outflow
while the gas moves parallel to the $z$-axis (panels 3-a, 3-b and 3-d).
This indicates that the dust decoupled from the gas is ejected from the
rotating outflow due to centrifugal force. Furthermore, some dust stream-lines do not
diverge but take a closed path, meaning that some of large dust is refluxed from the outflow to the envelope.
The ejection further decreases the amount of the dust in the outflow.
In addition, the ejected dust gathers around the outflow,
forming a U-shaped dust enhanced region, which is clearly visible in the dust-gas mass ratio map (panel 3-d).

At $t_*=1.1\times 10^4$ yr (panels 4-a to 4-d),
the dust with a size of $a_d \gtrsim 1 \mm$  is distributed over a large area of several 100 AU across (panel 4-c).
The stream-lines and velocity field of the dust clearly indicate that the dust
moves from the outflow back to the envelope (panels 4-a, b, d),
while the stream lines and velocity field of gas 
shows that the gas is released to the interstellar medium.
At this epoch, the gas and dust density distribution in the outflow are markedly different.
This is caused by the dust-gas decoupling in the outflow. The majority
of the large dust in the outflow have been refluxed to the envelope.

\subsection{``Ash-fall": reflux of large dust from outflow to the disk outer edge}
\begin{figure*}
  \includegraphics[width=170mm]{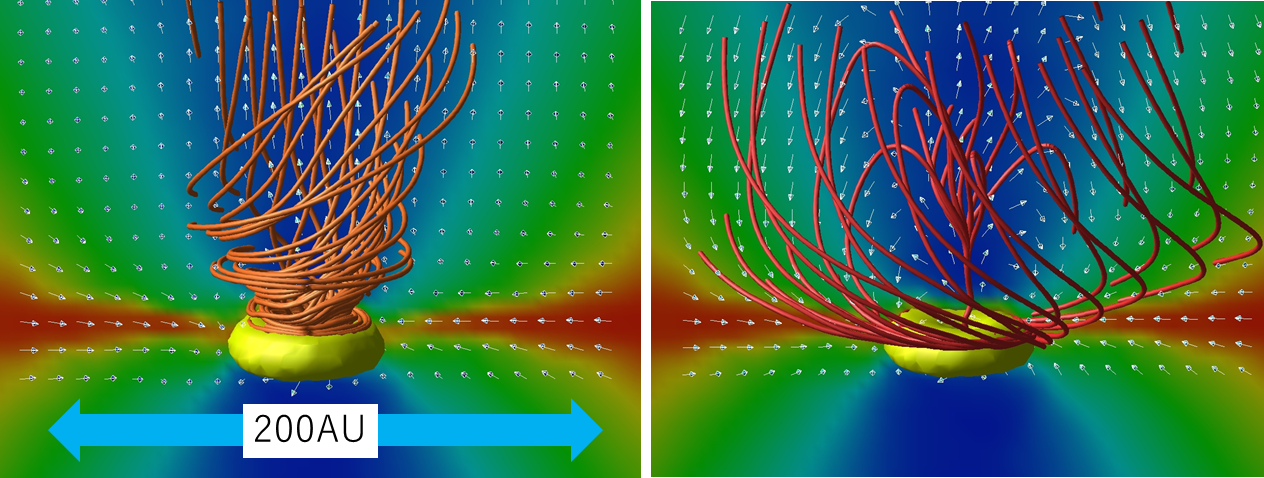}
\caption{
  Left and right panels show stream-lines of gas (orange) and dust (red) at $t_*=1.1\times 10^4$ yr, respectively.
  Arrows indicate the velocity field of either of the gas and dust on the $x$-$z$ plane.
  The yellow isosurfaces show the protoplanetary disk.
  Color scale shows the dust-to-gas mass ratio and is identical between left and right panels.
}
\label{fig2}
\end{figure*}

Figure \ref{fig2} shows three-dimensional stream-lines of the gas and dust at $t_*=1.1\times 10^4$ yr.
The gas stream-lines show that the gas is rotating and outflowing from the surface of the disk
and is blown away.
By contrast, the dust stream-lines indicate different dusty dynamics from that of the gas.
The dust also outflows from the inner part of the disk.
Then it immediately decouples from the gas in the outflow. Since the outflow rotates as shown in the stream line of gas,
the centrifugal force drives the dust towards the radial direction.
Gravity then pulls it back towards the disk and the grown dust accretes to the edge of the disk.

\begin{figure*}
  \includegraphics[width=170mm]{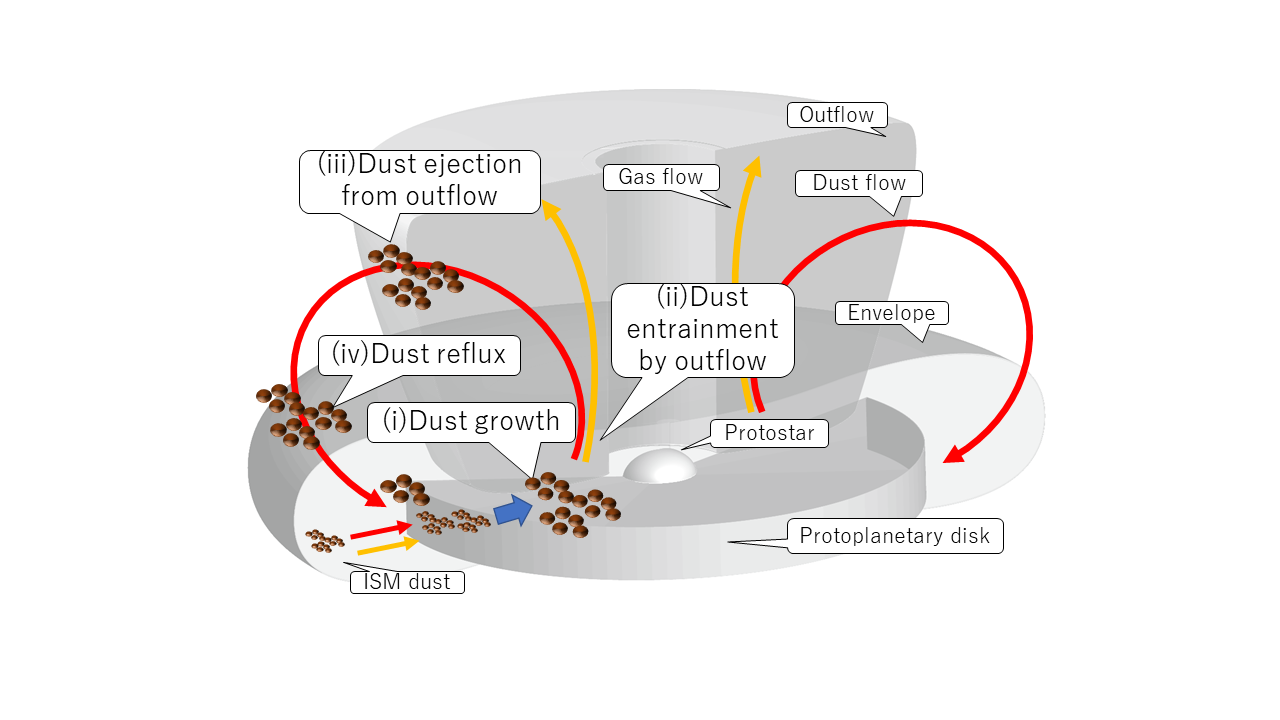}
\caption{
   The schematic drawing of overall dust evolution in "ash-fall'' phenomenon.
}
\label{schematic_fig2}
\end{figure*}

The overall process is schematically shown in the figure \ref{schematic_fig2}, and 
is very similar to ash-fall from volcanic eruptions where dust-gas mixture is ejected from the crater
and then the gas and dust decouple in the atmosphere, causing the dust (or ash) to fall selectively.
Hence we name this phenomenon as ``ash-fall"  phenomenon in protostar evolution.

\subsection{Amount of the refluxed dust}
\begin{figure*}
\includegraphics[angle=-90, width=100mm]{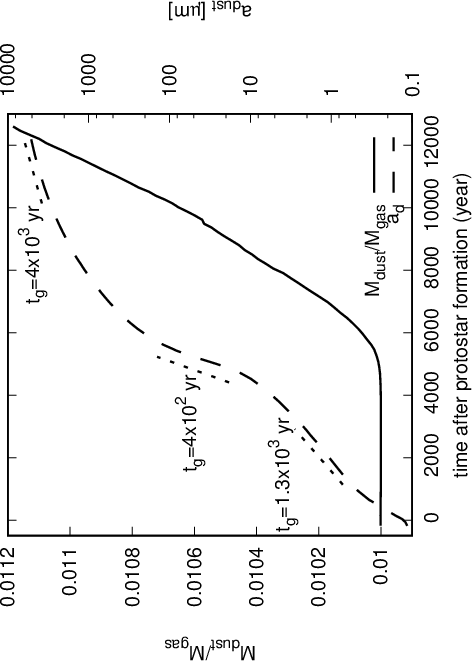}
\caption{
  Evolution of the (solid line) dust-to-gas mass ratio and (dashed line) mean dust size  in the protoplanetary disk as a function of time
  after the protostar formation.
  Dotted lines show $a_d\propto \exp(t/t_g)$ for growth timescales of $t_g=1.3\times 10^3,4 \times 10^2,4\times 10^3$ yr.
}
\label{fig3}
\end{figure*}

The reflux of the large dust due to the  ash-fall phenomenon increases
the amount of the dust in the accretion flow from the envelope
and thus the dust-to-gas mass ratio in the disk.
Figure \ref{fig3} shows the time evolution of the mean dust-to-gas mass ratio
of the disk $M_{d,disk}/M_{g,disk}$ and the mean dust size $\int \rho a_d dV/\int \rho dV$ 
in the disk, where the integration is performed within the disk.
Here, we define the disk as a region with the gas density $\rho_g>10^{-13} \gcm$.
We found that, in an early phase of  $t_*<5\times 10^3$ yr, where the mean dust size is $a_d\lesssim 10^2 \mum$,
the dust-to-gas mass ratio stays constant at $M_{\rm dust}/M_{\rm gas} =10^{-2}$, maintaining the initial value.
This fact implies that the amount of the refluxed dust is negligibly small, presumably because
the outflowing dust is well coupled with gas and is not ejected from the outflow.

Once dust has grown to a size of $a_d\gtrsim 10^2 \mum$ at $t_*\sim 4 \times 10^3 $ yr, the dust-to-gas mass ratio begins to increase
as a result of the increase of the amount of the dust in the accretion flow from the envelope due to the dust reflux.
The dust-to-gas mass ratio keep increasing and becomes $\epsilon \sim 0.01$ at the end epoch of the simulation ($t_*\sim 1.2 \times 10^4$ yr),
implying that $\sim 10$ \% of the dust in the disk is the refluxed large dust.

  Note that the increase of dust-to-gas mass ratio in the disk is
  a consequence of the dust growth in the disk and the dust reflux,
  and hence, in contrast to \citet{2020A&A...641A.112L},
  is not a consequence of the dust drift in the prestellar collapse phase.
  We start from the dust size with $0.1 \mum$ and, with this dust size,
  the dust and gas are essentially completely coupled and maintain the initial dust-to-gas
  mass ratio of $10^{-2}$ during prestellar collapse phase (see figure \ref{fig1}-d and \ref{fig3} at $t_*=0$).

  Note also that $a_d$ obtained in the single-size approximation represents the peak-mass dust size \citep{2016A&A...589A..15S} and the contribution of smaller dust
  on the dust density may be minor (at least with MRN like size distribution).
  Therefore, we expect $\sim 10$\% increase in dust-to-gas mass ratio
  in the disk shown in figure \ref{fig3} may be realized even by considering
  a more realistic dust size distribution.
  Future studies with dust size distribution are imperative to confirm this prediction.


\subsection{Enhancement of dust-to-gas mass ratio in the upper envelope and formation of U-shaped dust-enhanced region}
\begin{figure*}
\includegraphics[width=180mm]{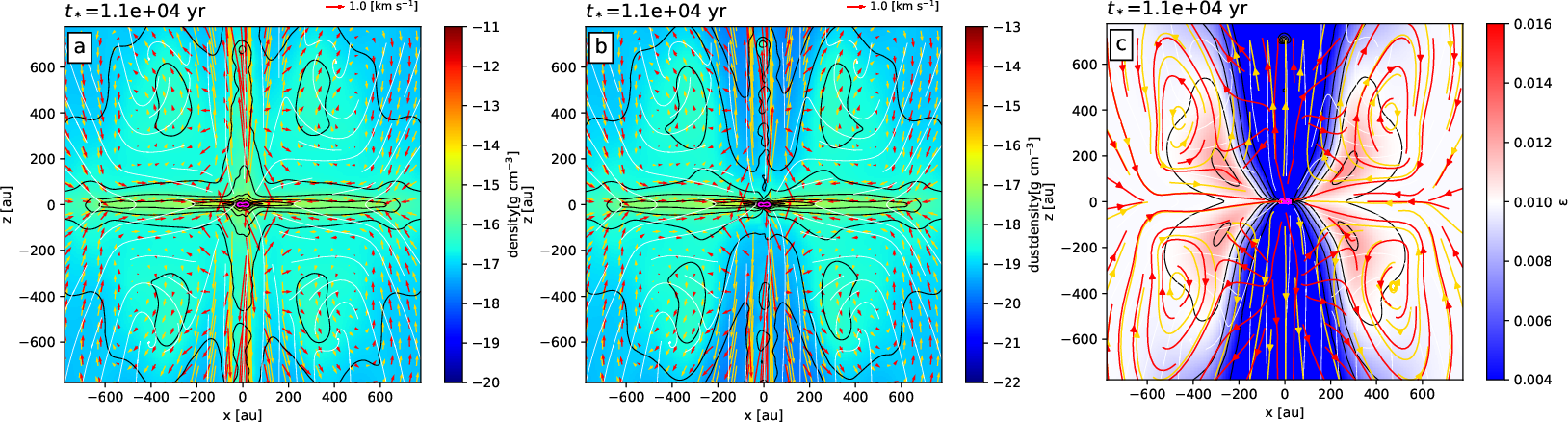}
\caption{
  Large-scale (1500 AU across) structure of the (panel a) gas density, (b) dust density, and (c) dust-to-gas mass ratio
  at $t_*=1.1 \times 10^4$ yr on $x$-$z$ plane as in figure \ref{fig1}.
  See the caption of figure \ref{fig1} for notation.
  The contour levels are (panel a) $\rho_g=10^{-20},10^{-19.5},\cdots,10^{-11}\gcm$, (b)$\rho_d=10^{-22},10^{-21.5},\cdots,10^{-13}\gcm$,
  and (c) $\epsilon=4\times 10^{-3},4\times 10^{-3},\cdots, 1.6 \times 10^{-2}$, respectively.
}
\label{fig4}
\end{figure*}

Figure \ref{fig4} shows the gas density, dust density, and dust-gas ratio for a 1500 AU scale at $t_*=1.1\times 10^4$ yr.
The gas density map shows that the gas is relatively dense inside the outflow
while the dust density is significantly depleted inside the outflow and is enriched around the outflow, forming a U-shaped dust enhancement.
In the dust-enhanced region, the dust-to-gas mass ratio is as high as 0.015, which is roughly a 50\% increase from that of the typical interstellar medium.

A similar U-shaped dust-enhanced region has been observed in some Class 0/I YSOs,
where the dust thermal emission was not detected, or much weaker than the surrounding regions, inside the outflow
\citep[such as L1527, HOPS136, B335, IRAS 15398-3559][]{2010ApJ...710.1786Y,2014ApJ...781..123F,2017ApJ...834..178Y,2018MNRAS.477.2760M,2018ApJ...861...91H}.
The reason of lack of detection of the dust thermal emission  inside the outflow while
it was detected in the surrounding regions, has been thought to be the low gas density in the outflow.
However, our simulation suggests that  dust cavities are formed due to the ejection of large dust from the outflow by centrifugal force.
Then, it implies that  the U-shaped dust enhancement is an indicator of growth of dust in the disk and outflow.

\section{Discussion}
\subsection{"ash-fall" phenomenon in protostar evolution}
In this paper, we propose a new type of dust dynamics  i.e., the ``ash-fall" phenomenon in protostar evolution,
in which the grown dust in the disk is blown away by outflows and is then refluxed on to the outer edge of the disk.
The process consists of the following four steps (see also the schematic drawing in figure \ref{schematic_fig2}):
  \begin{enumerate}
  \item dust growth in the disk where the density is sufficiently high and the dust growth timescale $t_{\rm growth}$ is as short as  $t_{\rm growth}\sim 10^3$ yr (see Appendix B and figure 4),
  \item large dust with $a_d>1 \mm$ can be entrained by the powerful magnetized outflow launched from the inner region of the disk,
  \item decoupling of the gas and dust in the outflow and dust ejection from the outflow due to centrifugal force, leading U-shaped dust-enhanced region around the outflow, 
  \item mixing of the grown dust into the envelope accretion flow and the reflux of the large dust to the disk outer edge.
  \end{enumerate}

  The ash-fall supplies a non-negligible amount of large dust to the outer-edge of the disk even at shortly after protostar formation ($t\sim 10^4$ yr) and keep increasing
  ($\sim 10$ \% of the total amount of the accreting dust from the envelope; see figure \ref{fig3}).


\subsection{Implications for planet formation}
The outer edge of a disk has a lower surface density than its inner region and
the Stokes number $\St$ is inversely proportional to the gas surface density $\Sigma_g$,
given by $\St=\pi \rho_{\rm mat} a_d/(2 \Sigma_g)$, where $\rho_{\rm mat}$ is the dust internal density.
Therefore, the Stokes number of dust increases when the dust is refluxed to the disk outer edge.
As a result, the dust of $\St<1$ at the inner region of the disk can have $\St>1$ when it is refluxed to the disk edge.


More quantitatively, we make the following estimates.
We assume the disk has an exponential tail \citep[as suggested from observations][]{2011ARA&A..49...67W}
and connected to envelope with accretion shock \citep[e.g., ][]{2017ApJ...839...47M}.
The envelope surface density at the connection with the edge of the disk
(at the upstream of accretion shock) is estimated as
$\Sigma_g=\dot{M}_{\rm gas}/(2 \pi r v_r)$ where $v_r =\sqrt{2 G M_{\rm star}/r}$.
Then the disk surface density at the edge of the disk
(at the downstream of accretion shock) is roughly given as
\begin{eqnarray}
\Sigma_g \sim \left(\frac{v_r}{c_s}\right)^2 \frac{\dot{M}_{\rm gas}}{2 \pi r v_r},
\end{eqnarray}
by assuming the shock is isothermal.
Hence the Stokes number at the disk outer edge is
\begin{eqnarray}
  \label{St_estimate}
  \St=25 \left( \frac{ r}{100 {\rm AU}}\right)^{1/2} \left( \frac{\rho_{\rm mat}}{2 \gcm}\right)   \left( \frac{a_{d}}{ 1 \cm}\right) \nonumber \\
 \left( \frac{M_{\rm star}}{ 0.3 \msun}\right)^{-1/2}\left( \frac{\dot{M}_{gas}}{ 10^{-7} \msunyear}\right)^{-1} \left( \frac{c_s^2}{ 350 \ms}\right)^{2},
\end{eqnarray}
where $\dot{M}_{gas}$ is the mass accretion rate.
Note that $\St$ can be even larger
by a factor of $(v_r/c_s)^2 (\sim 40$ with our parameters)  when the accretion shock
is adiabatic and density increase by the shock compression is small.
Note also that the downstream density also decreases when the shock is oblique.
This estimate suggests that the dust of $a_d \gtrsim 1 \cm$ can be already $\St \gtrsim 1$ when it is refluxed to disk edge.

As such, the ash-fall phenomenon can provide a pathway to circumvent the radial drift barrier
and may allow dust to grow to planetesimals in the outer region of disk.
Thus, the large dust could be the embryo of the planetesimal at the outer region of the disk.

 Note that the radial drift barrier is conventionally regarded as a problem in Class II/III YSOs.
 Actually, the radial drift universally occurs even in Class 0/I YSOs,
 once the rotationally supported protostellar disk is created and dust growth starts. 
 For example, \citet{2017ApJ...838..151T} showed that the dust growth
 and radial drift also happens in a disk of Class 0/I YSOs.
 In this sense, the occurrence of the radial drift barrier is not limited to Class II/III YSOs.

 On the other hand, some molecular outflows seem to last
 longer and remain even in Class II YSOs \citep[e.g.,][]{1995ApJ...452..736H}.
 Thus, although our simulation stops at $t \sim 10^4$ yr after protostar
 formation (i.e., investigating the evolution of Class 0/I YSOs), we can expect ash-fall mechanism also plays a role even in Class II YSOs
 (or possibly more important because Stokes number of refluxed dust at the disk edge
 increases as envelope accretion diminishes; see, equation (\ref{St_estimate})).

  In addition, a significant amount of large dust whose Stokes number is close to unity refluxed to disk 
  provides an ideal condition for the growth of the secular gravitational instability (SGI) \citep{2014ApJ...794...55T,2016AJ....152..184T,2018PASJ...70....3T,2020ApJ...900..182T}
  that are expected to be a powerful mechanism for creating ring and planetesimals in the outer region of the disk.
  Since SGI creates many axisymmetric dusty ring structures in the disk and may subsequently create planetesimals,
  a combination of the ash-fall and subsequent SGI can be an explanation for multiple ring-like structure observed in protoplanetary disks
  \citep{2015ApJ...808L...3A,2017A&A...600A..72F,2018ApJ...869L..41A,2018A&A...610A..24F,2018ApJ...869...17L}.
  It is interesting to note that the very first example of those disks, HL-Tau, is known to possess bipolar molecular outflow
  and collimated jet-like structures. The co-existence of outflows/jets and very ordered
  axisymmetric multiple ring-like structure in the disk has been puzzling us since its discovery in 2014.
  Now we are proposing that actually these two processes might be cooperating in such an object.

\subsection{Implication for observations of Class 0 YSOs}

  Another important implication of ash-fall is that the reflux of the large dust into the envelope of $\sim 1000$ AU  may explain the presence of grown dust in the envelope suggested by the observations.
  As shown in, e.g., \citet[][]{2009ApJ...696..841K} and \citet{2019A&A...632A...5G},
  the spectral index of dust opacity $\beta$ in the envelope of some Class 0 YSOs is lower than that of the ISM.
  We expect that this is explained by large dust supplied from the disk to the envelope by the outflow.


\subsection{Comparison with previous studies}

  Recently, \citet{2020A&A...641A.112L} conducted the non-ideal MHD simulation with dust fluids including dust-size distribution for the first time.
  Here we summarize the novel points of our work in comparison with \citet{2020A&A...641A.112L}.
  
  The essential points of our study are that we show, for the first time, with 3D non-ideal MHD
  simulation  with {\it dust growth}, that the chain of (i) to (iv) (see above)
  self-consistently occurs. We also suggest that this mechanism can be
  a theoretical breakthrough to overcome the radial drift barrier.

  On the other hand, \citet{2020A&A...641A.112L}, in the context of (i),
  discuss the importance of dust growth in the discussion section (section 7.5,
  "Caveat: coagulation and fragmentation during the collapse")
  but do not include dust growth in their simulations.
  We improve this point by actually solving the dust growth in the simulation. 
  In the context of (ii), they show that dust size with $\sim 100 \mum$ can be entrained by outflow.
  On the other hand, we show that much larger dust of $a_d>1 \mm$ (which has more than ten times longer stopping time) can be entrained.
  In the context of (iii), from their results, whether the dust decoupling and dust ejection from outflow occur is not clear.
  They show that the dust is enriched (rather depleted) in outflow, suggesting that the dust and gas are relatively well coupled.
  This may be reasonable because the dust stopping time in outflow in their simulations is more than ten times smaller than in ours.
  Note, however, that they set a numerical upper limit of $1 \kms$ on the dust-gas relative velocity,
  which is smaller than the relative velocity realized in our simulaiton (see arrows of figure 4-a or 4-b).
  We think this treatment may restrict the range of relative motion between gas and dust and affect the dynamics of dust grains such as ejection. 
  We improved this point by explicitly solving the time dependence of the relative velocity without velocity limit.
  In the context of (iv),  the reflux to the disk outer edge, seems to be missing in their results or presentation.

Finally, we briefly comment on the relation of our ash-fall phenomenon to the long-standing problem
of chondrule and calcium-aluminum-rich inclusions (CAIs) formation.
By showing that dust reflux can happen, our simulation supports
the formation model of CAIs and chondrule with a bipolar outflow
\citep[e.g.,][]{2001ApJ...548.1029S,2019AJ....158...55H}, where the reflux
is analytically discussed to occur from the inner to outer parts of a disk.
New insights into the formation process of CAIs and chondrule will be gained with future studies
of two-fluid radiation magnetohydrodynamics simulation of the dusty gas with the resolution to the vicinity of the central protostar.




\section*{Acknowledgments}
We thank Dr. Satoshi Okuzumi for their comments.
The computations were performed on the Cray XC50 system at CfCA of NAOJ.
This work is supported by JSPS KAKENHI  grant number 18H05437, 18K13581, 18K03703.


\appendix

\section{Dust relative velocity}
For the turbulent induced dust relative velocity, we adopt the form presented by \citet{2007A&A...466..413O},
\begin{eqnarray}
\Delta v_{\rm turb}=
\begin{cases}
  \frac{\delta v_{\rm Kol}}{t_{\rm Kol}} (t_{\rm stop, 1}-t_{\rm stop, 2}) & (t_{\rm stop, 1}<t_{\rm Kol})\\
  1.5 \delta v_L \sqrt{\frac{t_{\rm stop, 1}}{t_L}} & (t_{\rm Kol}< t_{\rm stop, 1}<t_L) \\
  \delta v_L \sqrt{\frac{1}{1+t_{\rm stop, 1}/t_L}+\frac{1}{1+t_{\rm stop, 2}/t_L}} & (t_L< t_{\rm stop, 1})
\end{cases}
\end{eqnarray}
where  $\delta v_{\rm Kol}=Re_L^{-1/4}$ and  $ \delta v_L, ~ t_{\rm Kol}=Re_L^{-1/2} t_L$ are the
eddy velocity and eddy turn-over timescale at dissipation scale and $Re_L=L v_L/\nu$ is the Reynolds number.
We set $t_{\rm stop, 1}=t_{\rm stop}(a_d)$ and  $t_{\rm stop, 2}=1/2~ t_{\rm stop}(a_d)$ referring to \citet{2016A&A...589A..15S}.
For the sub-grid model of turbulence, we consider ``$\alpha$ turbulent model ", in which 
we choose
\begin{eqnarray}
  \delta v_L=\sqrt{\alpha_{\rm turb}} c_s, \\
  \label{tL_eq}
  t_L=\frac{c_s}{a_g}, \\
  L=\delta v_L t_L,
\end{eqnarray}
where $\alpha_{\rm turb}=2 \times 10^{-3}$ is the dimensionless parameter which determines the strength
of the sub-grid turbulence and  $a_g$ is the gravitational acceleration.
The fluctuating velocity at the largest scale corresponds to $\delta v_L =\sqrt{\alpha_{\rm turb}} c_s \sim 0.05 c_s$, which is a conservative value
(for example, $\delta v_L\sim c_s$ ($\alpha_{\rm turb} =1$) or $\delta v_L\sim 0.1 c_s$ ($\alpha_{\rm turb} =10^{-2}$) is often assumed in envelope or disk in the
studies of dust growth), and we do not artificially accelerate dust growth.
The reason of our choice  $t_L$ (Equation (\ref{tL_eq})) is in a word,
to model the turbulence in both  envelope and disk in a common way.
The former structure is determined by the self-gravity whereas the latter by the gravity of the central protostar.
In our modelling,  the length scale and largest eddy-turn over timescale in the case where the self-gravity determines the structure are
\begin{eqnarray}
  a_g \sim \frac{G M(L)}{L^2}\sim G \rho L, \nonumber \\
  t_L=\frac{c_s}{a_g} \sim t_{\rm ff},
\end{eqnarray}
when we assume $\alpha_{\rm turb}=1$, where $G$ is the gravitational constant and  $M(L)$ is the enclosed mass within the length scale $L$.
Thus, the situation that our model assumes corresponds to the condition of  $t_L\sim t_{\rm ff}$ and $L\sim \lambda_J$ which have been
used in modelling of dust growth in the cloud core and envelope in past studies \citep{2009A&A...502..845O,2013MNRAS.434L..70H}.

By contrast, in the case where the gravity of the central protostar determines the structure (which primarily corresponds to disk),
the length scale and largest eddy-turn over timescale are
\begin{eqnarray}
  a_g\sim \frac{G M_*}{r^2}, \nonumber \\
  t_L= \frac{c_s}{a_g} \sim \frac{H}{r}\Omega^{-1},
\end{eqnarray}
where $M_*$ is the protostar mass.
Thus, the situation that our model assumes corresponds to the condition of $t_L \sim \Omega^{-1} \frac{H}{r}$.
Note that the turbulence induced by magneto-rotational instability (MRI) may have $t_L \sim \Omega^{-1}$ 
\citep{2006A&A...452..751F} which is often used in the modelling of the turbulence in the disk.
Thus, our model may introduce a larger relative velocity by factor of two or three
than that induced by MRI ($H/r\gtrsim 0.1$ for $r \gtrsim 1$ AU in our simulation).
The turbulence induced by vertical shear instability (VSI), by contrast, may have 
$t_L \sim 10^{-1} \Omega^{-1}$ \citep{2016A&A...594A..57S}.
Thus, $t_L$ of our sub-grid turbulence model falls in between those in MRI and VSI in the disk.
In practice, the uncertainty is absorbed in the parameter $\alpha_{\rm turb}$.

The relative velocity induced by Brownian motion is assumed to be
\begin{eqnarray}
  \Delta v_B=\sqrt{\frac{8 m_g}{\pi m_d}}c_s.
\end{eqnarray}


\section{Dust growth timescale in the envelope and disk}

First, we estimate $\tgrowth$ of envelope.
By assuming that the relative velocity among the dust is determined by turbulence and
$t_{\rm kol}<\tstop<t_L$,  $t_L=t_{\rm ff}$, and  $\tstop=\rho_{\rm mat} a_d/(v_{\rm therm} \rho_g)$,
$\tgrowth$ in envelope is given as \citep{2007A&A...466..413O},
\begin{eqnarray}
  \tgrowth
  &=& 1.7 \times 10^4  \dv_{L,190 \ms}^{-1} \rho_{\rm mat, 2 \gcm}^{1/2} a_{d, 1 \mum}^{1/2} \nonumber \\
  & &f_{0.01}^{-1} \rho_{g,10^{-16} \gcm}^{-3/4} c_{s,190 \ms}^{1/2}  ~{\rm year},
\end{eqnarray}
where we assume $\rho_d=f \rho_g$ and $f=10^{-2}$ is the dust-to-gas mass ratio.
For each quantity, $f_{X}$ means $f_{X}=(\frac{f}{X})$.
The ratio of the growth timescale to the free-fall timescale is given as
\begin{eqnarray}
  \frac{\tgrowth}{t_{\rm ff}}&=&2.5
  \dv_{L,190 \ms}^{-1} \rho_{\rm mat, 2 \gcm}^{1/2} a_{d, 1 \mum}^{1/2} f_{0.01}^{-1} \nonumber \\
  & & \rho_{g,10^{-16} \gcm}^{-1/4} c_{s,190 \ms}^{1/2}.
\end{eqnarray}
This highlight the difficulty of dust growth in the cloud core and envelope and 
dust may not grow to $\gtrsim 1 \mum$ in the envelope.

Next, we estimate $\tgrowth$ of disk.
By assuming that the relative velocity among the dust is determined by turbulence and
$t_{\rm kol}<\tstop<t_L$,  $\dv_L=\sqrt{\alpha c_s^2}$, and $t_L=\Omega^{-1}$
\begin{eqnarray}
  \tgrowth &=& 2.6 \times 10^3  \alpha_{10^{-2}}^{-1/2} \rho_{\rm mat,2 \gcm}^{1/2}  a_{d,  1 \mm}^{1/2} f_{0.01} ^{-1}  \\
  & &\rho_{g,10^{-12} \gcm}^{-1/2}c_{s,190 \ms}^{-1/2}  M_{*, 0.1 \msun}^{-1/4} r_{10 {\rm AU}}^ {3/4} ~{\rm year}. \nonumber 
\end{eqnarray}
In contrast to the dust growth in the envelope,
the dust can quickly grows to the order of $\mm$
in disk thanks to its large density even in Class 0/I YSOs.

\bibliography{article}

\end{document}